\documentclass[12pt,a4paper]{article}

\usepackage{ifthen} 
\newboolean{pdflatex}
\setboolean{pdflatex}{true} 

\newboolean{articletitles}
\setboolean{articletitles}{true} 

\newboolean{uprightparticles}
\setboolean{uprightparticles}{false} 

\newboolean{inbibliography}
\setboolean{inbibliography}{false} 

\usepackage[top=1in, bottom=1.25in, left=1in, right=1in]{geometry}
\usepackage{microtype}
\usepackage{lineno}  
\usepackage{xspace} 
\usepackage{caption} 

\usepackage{graphicx}  
\usepackage{color}
\usepackage{colortbl}
\graphicspath{{./figs/}} 
\usepackage{amsmath} 
\usepackage{amssymb}
\usepackage{amsfonts}
\usepackage{upgreek} 
\newcommand*\patchAmsMathEnvironmentForLineno[1]{%
\expandafter\let\csname old#1\expandafter\endcsname\csname #1\endcsname
\expandafter\let\csname oldend#1\expandafter\endcsname\csname
end#1\endcsname
 \renewenvironment{#1}%
   {\linenomath\csname old#1\endcsname}%
   {\csname oldend#1\endcsname\endlinenomath}%
}
\newcommand*\patchBothAmsMathEnvironmentsForLineno[1]{%
  \patchAmsMathEnvironmentForLineno{#1}%
  \patchAmsMathEnvironmentForLineno{#1*}%
}
\AtBeginDocument{%
\patchBothAmsMathEnvironmentsForLineno{equation}%
\patchBothAmsMathEnvironmentsForLineno{align}%
\patchBothAmsMathEnvironmentsForLineno{flalign}%
\patchBothAmsMathEnvironmentsForLineno{alignat}%
\patchBothAmsMathEnvironmentsForLineno{gather}%
\patchBothAmsMathEnvironmentsForLineno{multline}%
\patchBothAmsMathEnvironmentsForLineno{eqnarray}%
}
\usepackage{hyperref}    
\usepackage[all]{hypcap} 

\def\lhcb   {\mbox{LHCb}\xspace}

\def\babar  {\mbox{BaBar}\xspace}
\def\belle  {\mbox{Belle}\xspace}

\def\CP  {{\ensuremath{C\!P}}\xspace}

\newcommand{\ie}{\mbox{\itshape i.e.}\xspace}

\usepackage{xspace}
\usepackage{upgreek}

\ifthenelse{\boolean{uprightparticles}}%
{

 \def\Ppi         {\ensuremath{\uppi}\xspace}

 \def\PDelta      {\ensuremath{\Delta}\xspace}
 \def\PXi      {\ensuremath{\Xi}\xspace}
 \def\PLambda      {\ensuremath{\Lambda}\xspace}
 \def\PSigma      {\ensuremath{\Sigma}\xspace}
 \def\POmega      {\ensuremath{\Omega}\xspace}
 \def\PUpsilon      {\ensuremath{\Upsilon}\xspace}

 \def\PB      {\ensuremath{\mathrm{B}}\xspace}

 \def\PK      {\ensuremath{\mathrm{K}}\xspace}

 \def\Pb      {\ensuremath{\mathrm{b}}\xspace}
 \def\Pc      {\ensuremath{\mathrm{c}}\xspace}

 \def\Pi      {\ensuremath{\mathrm{i}}\xspace}

 \def\Pn      {\ensuremath{\mathrm{n}}\xspace}
 
 \def\Pp      {\ensuremath{\mathrm{p}}\xspace}

 \def\Ps      {\ensuremath{\mathrm{s}}\xspace}

}
{

 \def\Ppi         {\ensuremath{\pi}\xspace}

 \mathchardef\PDelta="7101
 \mathchardef\PXi="7104
 \mathchardef\PLambda="7103
 \mathchardef\PSigma="7106
 \mathchardef\POmega="710A
 \mathchardef\PUpsilon="7107
 
 \def\PB      {\ensuremath{B}\xspace}

 \def\PK      {\ensuremath{K}\xspace}

 \def\Pb      {\ensuremath{b}\xspace}
 \def\Pc      {\ensuremath{c}\xspace}

 \def\Pi      {\ensuremath{i}\xspace}

 \def\Pn      {\ensuremath{n}\xspace}
 
 \def\Pp      {\ensuremath{p}\xspace}

 \def\Ps      {\ensuremath{s}\xspace}

}

\def\squark    {{\ensuremath{\Ps}}\xspace}

\def\cquark    {{\ensuremath{\Pc}}\xspace}
\def\cquarkbar {{\ensuremath{\overline \cquark}}\xspace}

\def\bquark    {{\ensuremath{\Pb}}\xspace}

\def\pion   {{\ensuremath{\Ppi}}\xspace}

\def\pim    {{\ensuremath{\pion^-}}\xspace}

\def\kaon   {{\ensuremath{\PK}}\xspace}
\def\Kbar    {{\kern 0.2em\overline{\kern -0.2em \PK}{}}\xspace}

\def\KorKbar {\kern 0.18em\optbar{\kern -0.18em K}{}\xspace}

\def\Kp      {{\ensuremath{\kaon^+}}\xspace}

\def\KS      {{\ensuremath{\kaon^0_{\mathrm{ \scriptscriptstyle S}}}}\xspace}

\def\B {{\ensuremath{\PB}}\xspace}

\def\Bds{{\ensuremath{\B^0_{\kern -0.1em{\scriptscriptstyle (}\kern -0.05em\squark\kern -0.03em{\scriptscriptstyle )}}}}\xspace}

\def\Bbar    {{\ensuremath{\kern 0.18em\overline{\kern -0.18em \PB}{}}}\xspace}

\def\BorBbar    {\kern 0.18em\optbar{\kern -0.18em B}{}\xspace}

\def\Bd      {{\ensuremath{\B^0}}\xspace}

\def\Bs      {{\ensuremath{\B^0_\squark}}\xspace}

\def\PLambda {\ensuremath{\Lambda}\xspace}
\def\proton      {{\ensuremath{\Pp}}\xspace}
\def\antiproton  {{\ensuremath{\overline \proton}}\xspace}
\def\neutron     {{\ensuremath{\Pn}}\xspace}

\def\Lz          {{\ensuremath{\PLambda}}\xspace}
\def\Lbar        {{\ensuremath{\kern 0.1em\overline{\kern -0.1em\PLambda}}}\xspace}
\def\Xires       {{\ensuremath{\PXi}}\xspace}

\def\Xim         {{\ensuremath{\Xires^-}}\xspace}

\def\Lb          {{\ensuremath{\Lz^0_\bquark}}\xspace}
\def\Lbbar       {{\ensuremath{\Lbar{}^0_\bquark}}\xspace}
\def\Xibz        {{\ensuremath{\Xires^0_\bquark}}\xspace}
\def\Xibm        {{\ensuremath{\Xires^-_\bquark}}\xspace}
\def\Omegares    {{\ensuremath{\POmega}}\xspace}

\def\Omegab      {{\ensuremath{\Omegares^-_\bquark}}\xspace}

\newcommand{\decay}[2]{\ensuremath{#1\!\to #2}\xspace}       

\newcommand{\LPPi}{\texorpdfstring{\decay{\Lz}{\proton \pim}}{}}

\newcommand{\LbPPbarN}{\texorpdfstring{\decay{\Lb}{\proton \antiproton \neutron}}{}}
\newcommand{\LbLPPbar}{\texorpdfstring{\decay{\Lb}{\Lz \proton \antiproton}}{}}
\newcommand{\LbLLbarL}{\texorpdfstring{\decay{\Lb}{\Lz \Lbar \Lz}}{}}
\newcommand{\XibzLPPbar}{\texorpdfstring{\decay{\Xibz}{\Lz \proton \antiproton}}{}}
\newcommand{\XibzLLbarL}{\texorpdfstring{\decay{\Xibz}{\Lz \Lbar \Lz}}{}}
\newcommand{\XibmLLPbar}{\texorpdfstring{\decay{\Xibm}{\Lz \Lz \antiproton}}{}}

\newcommand{\OmegabXiPPbar}{\texorpdfstring{\decay{\Omegab}{\Xim \proton \antiproton}}{}}

\usepackage{cite} 
\usepackage{mciteplus}

\usepackage{longtable} 

\begin{document}

\renewcommand{\thefootnote}{\fnsymbol{footnote}}
\setcounter{footnote}{1}
\begin{titlepage}
\pagenumbering{roman}

\noindent
\begin{tabular*}{\linewidth}{lc@{\extracolsep{\fill}}r@{\extracolsep{0pt}}}
 & & \\
\end{tabular*}

\vspace*{3.0cm}

{\normalfont\bfseries\boldmath\huge
\begin{center}
Baryon decays\\
to purely baryonic final states
\end{center}
}

\vspace*{2.0cm}

\begin{center}
{\large Y.K. Hsiao$^{1}$, C.Q. Geng$^{1,2}$ and Eduardo Rodrigues$^{3,*}$}\\
\vspace*{0.2cm}
{\it
$^{1}$School of Physics and Information Engineering,\\Shanxi Normal University, Linfen 041004, China\\
$^{2}$Department of Physics, National Tsing Hua University, Hsinchu, Taiwan 300\\
$^{3}$Department of Physics, University of Cincinnati, Cincinnati, Ohio 45221, USA
$^{*}$Correspondence to eduardo.rodrigues@cern.ch
}
\end{center}

\vspace{\fill}

\begin{abstract}
\noindent
The \lhcb collaboration has presented first experimental evidence that spin-carrying matter
and antimatter differ. The study looked at four-body decays of the \Lb baryon.
Differences in the behaviour of matter and antimatter are associated with the
non-invariance of fundamental interactions under the combined charge-conjugation and
parity transformations, known as \CP violation.
We discuss purely baryonic decay processes, \ie decay processes involving only
spin-carrying particles. They are yet unexplored elementary processes.
Their study opens a new chapter of flavour physics in the route towards a better
understanding of \CP violation. It may help us
understand the observed matter and antimatter asymmetry of the Universe.
\end{abstract}

\vspace{\fill}
\end{titlepage}
\newpage
\setcounter{page}{2}
\mbox{~}
\clearpage

\renewcommand{\thefootnote}{\arabic{footnote}}
\setcounter{footnote}{0}


\pagestyle{plain} 
\setcounter{page}{1}
\pagenumbering{arabic}


\section*{Introduction}
\label{sec:introduction}
The \lhcb collaboration has presented first experimental evidence that
spin-carrying matter and antimatter differ~\cite{LHCb-PAPER-2016-030}.
Differences in the behaviour of matter and antimatter are associated with the
non-invariance of fundamental interactions under the combined
charge-conjugation and parity transformations, known as \CP violation.
Up until then, \CP violation had only been verified experimentally
with spin-zero mesons;
a brief historical review is given in Ref.~\cite{LHCb-PAPER-2016-030}.
As pointed out recently, the \lhcb measurement marks a first step into
unexplored territory~\cite{Durieux:2017nps}. It is of the utmost importance to
confirm the \lhcb result with higher statistical significance, analysing the
larger data samples now available from the second run of the
Large Hadron Collider at CERN.
Furthermore, numerous other decays of beauty baryons should be studied,
to establish a diverse set of observations, thereby improving our understanding
of \CP violation. Diversity of results comes in two flavours, namely from
the study of a variety of different systems, and via measurements of several
physical quantities sensitive to \CP violation.

\CP violation has far-reaching importance, being a crucial ingredient for the
generation of the observed matter-antimatter asymmetry in the Universe.
Unfortunately, our current theory and models can only explain a matter-antimatter asymmetry
at least ten orders of magnitude smaller than the one observed.
Additional sources of \CP violation, yet to be discovered, are likely to explain the discrepancy.
New sources of \CP violation may be seen again in the quark sector or in a different sector of the theory.
Since the visible Universe is made of spin-carrying particles such as the proton and the neutron,
it seems natural to study purely baryonic decay processes, \ie decay processes involving only
spin-carrying particles. Any \CP violating effects may have a more direct correspondence
to the long-standing puzzle of the matter-antimatter asymmetry.
These yet unexplored elementary processes may hold key information in much the same way
that the study of \CP violation with $B$ mesons provided a more comprehensive understanding
of \CP violation once it got established in the decay of neutral kaons.
Purely baryonic decay processes can exhibit a rich spin structure and provide complementary information
to that obtained so far with mesonic decays or final states.
For example, decays of baryons with spin of 1/2 or 3/2 can be used to construct
time-reversal violating observables, which provide other tests of \CP violation.

We discuss in this letter the study of purely baryonic decay processes.
For each beauty baryon we present the most promising decay mode to look for,
taking into account experimental constraints. Theoretical predictions are provided
for some decay branching fractions and, in some cases, for the \CP violating asymmetries.

\section*{Results}
\label{sec:results}
Elementary decay processes exclusively involving baryons are only kinematically allowed
with beauty baryons. These purely baryonic decays require at least three final-state
particles in order to fullfil the empirical law of baryon number
conservation~\cite{Geng:2016drz}.
The ``lowest-ground'' process is \LbPPbarN, discussed in Ref.~\cite{Geng:2016drz}.
We here focus our attention on the final states that are easiest to
reconstruct experimentally, in full, having in mind that the \lhcb collaboration
is the only running experiment capable of performing the search for these processes.
The lowest-ground beauty baryons of interest are the \Lb,
the isospin doublet \Xibz and \Xibm, and the \Omegab. The isotriplet $\Sigma_b$ baryons
decay strongly, hence this family is of little interest in the study of \CP violation
in weak decay processes.

The decay \LbLPPbar is a fully reconstructible final state. It does involve the
reconstruction of a long-lived particle, the \Lz baryon.
In \lhcb, long-lived particles are reconstructed with lower efficiencies than single
charged hadrons. Typically, an order of magnitude in selection efficiency is lost
for the presence of any single fully reconstructible long-lived particle in the final state
such as \Lz or \KS,
compared to the selection efficiency of reconstructing a charged
hadron. Still, the final state $\Lz \proton \antiproton$ seems the best way to observe
for the first time a fully baryonic final state of the \Lb baryon.

The \Xibz baryon can also decay to the $\Lz \proton \antiproton$ final state.
This decay is the most promising mode to observe a purely baryonic decay of the
\Xibz baryon. Indeed, moving up in complexity of reconstruction, both \Lb and \Xibz
can decay to the $\Lz \Lbar \Lz$ final state. This final state is unique in its own
right, and in particular provides a natural ground in which to study
the relatively poorly known decay modes of the charmonium $\cquark\cquarkbar$ resonances
to the $\Lz \Lbar$ final state.
The reconstruction efficiency of three long-lived \Lz baryons will unfortunately
be very low, which makes the decay modes \LbLLbarL and \XibzLLbarL out of reach until
the \lhcb experiment is upgraded for the years 2020s.

The search for purely baryonic decays of the \Xibm baryon is easiest performed looking
for the mode \XibmLLPbar. The reconstruction efficiency will be low owing to the need to
reconstruct two long-lived \Lz baryons.

The observation of a purely baryonic decay of the \Omegab will require large samples
yet to be collected by an upgraded \lhcb experiment. Its observation is presently
out of reach. On the one hand, the production rate of \Omegab is rather small compared
to the production of \Lb baryons. On the other hand,
the simplest decay mode is \OmegabXiPPbar, which involves the cascade \Xim in the
final state and hence the decay chain of two long-lived particles, as the \Xim baryon
is typically reconstructed in the $\Lz \pim$ final state. The resulting efficiency in the
reconstruction of the full decay chain is very low.

\subsection*{Branching fractions}
As mentioned above, purely baryonic decay processes were first considered in
Ref.~\cite{Geng:2016drz}, which focused attention on the simplest decay involving
the lightest possible baryons, \LbPPbarN. Its branching fraction is predicted to be
${\cal B}(\LbPPbarN) = (2.0^{+0.3}_{-0.2})\times 10^{-6}$~\cite{Geng:2016drz}.

The decays \LbLPPbar and \XibzLPPbar should be the easiest purely baryonic decay
processes to observe experimentally. Their branching fractions are predicted to be
$( 3.2 ^{+0.8}_{-0.3} \pm 0.4 \pm 0.7) \times 10^{-6}$ and
$(1.4\pm 0.1\pm 0.1\pm 0.4)\times 10^{-7}$,
where the uncertainties arise from non-factorisable effects, CKM matrix elements,
and hadronic form factors, respectively.

\subsection*{\CP asymmetries}
The study of triple-product correlations (TPCs) in three-body decays is handicaped
by the fact that the definitions of these TPCs involve the spin of one of the
final-state particles.
Such an issue does not happen in four-body decays, where TPCs depend only on the momenta
of the final-state particles.
The issue can nevertheless be overcome in specific cases, when dealing with so-called
self-tagging decay modes. The decay mode \LbLPPbar is such a decay. The charge of the
proton from the \LPPi decay automatically determines whether the decay is that of the \Lb baryon
or its \Lbbar antiparticle.

The direct \CP asymmetry of the \LbLPPbar decay is predicted to be
$( 3.4 \pm 0.1 \pm 0.1 \pm 1.0 ) \%$.
Similarly, the direct \CP asymmetry of the \XibzLPPbar decay is predicted to be
$(-13.0\pm 0.5\pm 1.5\pm 1.1)\%$.
Here, the first uncertainties account for non-factorisable effects, the second reflect the
experimental knowledge of the CKM matrix elements, and the third
correspond to those on the hadronic form factors (see Methods for a discussion of the latter).
The relatively large direct \CP asymmetry predicted for the \XibzLPPbar decay mode
makes it especially interesting from an experimental point of view.

\subsection*{Baryon-antibaryon enhancement near threshold}
Many $B$-meson decays to baryonic final states present a characteristic enhancement at (production)
threshold in the baryon-antibaryon mass spectrum of multi-body
decays~\cite{Hou:2000bz,Bevan:2014iga,LHCb-PAPER-2017-012,LHCb-PAPER-2017-005},
a fact that is still not fully understood.
Such enhancements are not observed in mesonic final states.
This same baryon-antibaryon enhancement near threshold is expected to be present in
the decays of $b$ baryons too. It awaits experimental confirmation.

\begin{figure}[t!]
\centering
\includegraphics[width=5.30in]{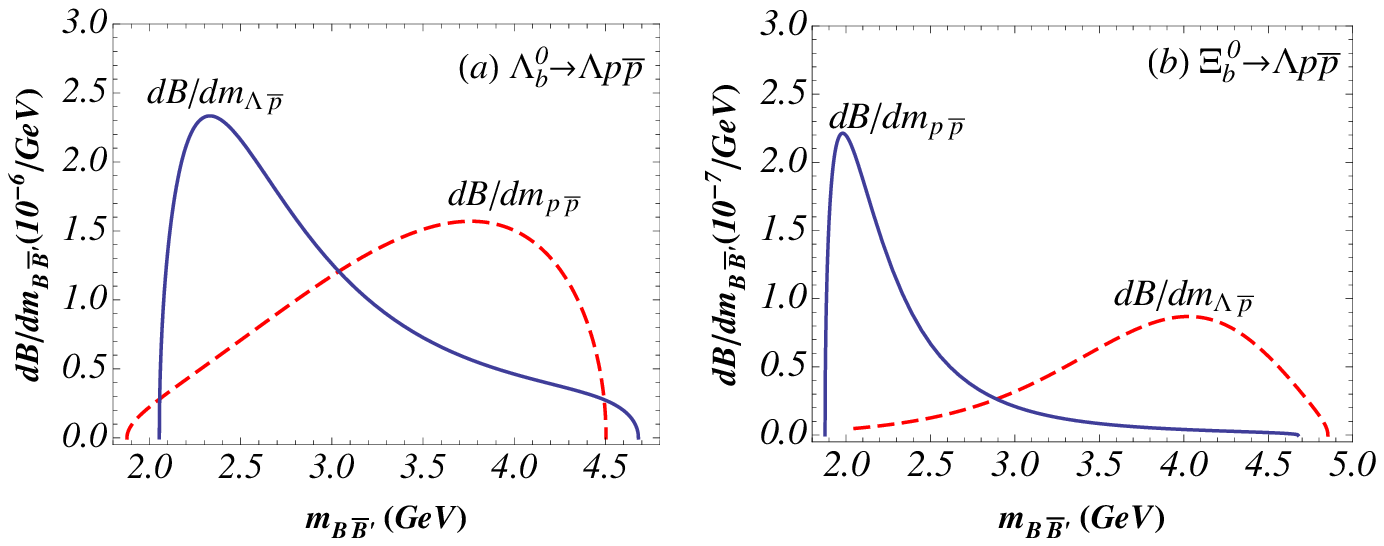}
\caption{The dibaryon invariant mass spectra for the
(a) \LbLPPbar and (b) \XibzLPPbar decays.}\label{mBB}
\end{figure}

Because of the participating Feynman diagrams, the \LbLPPbar and \XibzLPPbar decay modes
are expected not to exhibit a threshold enhancement in the same baryon-antibaryon system.
A threshold enhancement in $\Lz \antiproton$ is expected for the \LbLPPbar decay whereas
it is the invariant mass of the $\proton \antiproton$ system that is expected to
peak near threshold in the case of the \XibzLPPbar decay.
The expected dibaryon invariant mass spectra are displayed in Figure~\ref{mBB}.
These are clear signatures of the underlying QCD phenomenological framework used.

\section*{Discussion}
\label{sec:discussion}
The study of hadronic decays of $b$ hadrons has proved to be a rich playground
for a better understanding of \CP violation and for searches of manifestations of physics
beyond the Standard Model. The study of charmless decays, in particular,
has provided a wealth of crucial results and milestones in flavour physics,
notably the discovery of direct \CP violation in the $\Bd \to \Kp \pim$ decay~\cite{Aubert:2004qm,Chao:2004jy},
the first observation of \CP violation in the \Bs-meson system~\cite{LHCb-PAPER-2013-018},
and first evidence for \CP violation in the decay of a baryon,
\ie, in the decay of a spin-carrying particle~\cite{LHCb-PAPER-2016-030}.
Charmless decays, namely decays to final states with no charm flavour content,
typically involve flavour charged ($b \to u$) and neutral ($b \to s$ and $b \to d$)
transitions, which are suppressed with respect to the favoured $b \to c$ transition
to open-charm final states.

In the years to come, the \lhcb collaboration is the only running experiment
capable of studying beauty baryons. We urge the collaboration to expand
its presently ongoing programme of studies of $b$-hadron decays and to investigate
purely baryonic decays, which, for the first time, would allow a study
of \CP violation in decay processes involving only spin-carrying particles.
These yet unexplored elementary processes may hold key information towards a better
understanding of the \CP violating phenomena that are needed in order to explain
the observed matter and antimatter asymmetry of the Universe.

The decay modes \LbLPPbar and \XibzLPPbar are the most promising candidates
for the first observation of decay processes exclusively involving spin-carrying particles.
For what concerns \CP violation, although the current sensitivity of the \lhcb experiment
is unlikely to reach the level predicted in the Standard Model, it is still worthy to
explore the \CP violating asymmetries of fully reconstructed baryonic decays
as they could be large in models of physics beyond the Standard Model.

\section*{Methods}
\label{sec:methods}
\begin{figure}[tbhp]
\centering
\includegraphics[width=5.0in]{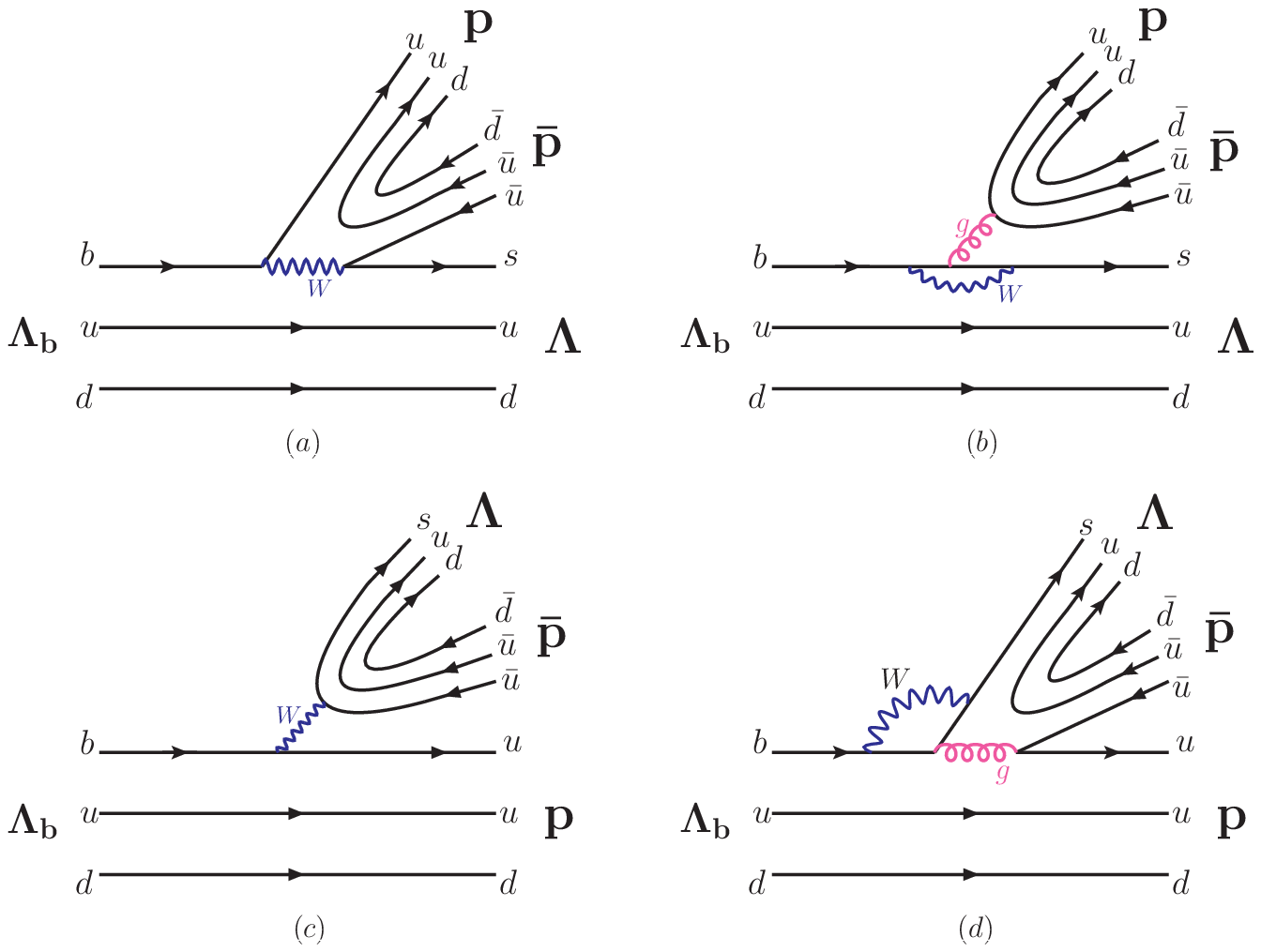}
\caption{Feynman diagrams describing the purely baryonic decay \LbLPPbar.}
\label{diagrams}
\end{figure}

Figure~\ref{diagrams} displays the dominant Feynman diagrams describing
the purely baryonic decays
${\bf B_b}\to {\bf B_1\bar B_2 B_3}$ (${\bf B}$ denotes a baryon),
which proceed with a $\bf B_b\to B_3$ transition
and a $\bf B_1\bar B_2$-pair production. The decay mode \LbLPPbar is taken as an example.
Similar diagrams can be drawn for the \XibzLPPbar decay.

According to Figure~\ref{diagrams},
the typical amplitude combines two matrix elements:
${\cal A}({\bf B_b}\to {\bf B_1\bar B_2 B_3})\sim
\langle {\bf B_1\bar B_2}|(\bar q_1 q_2)|0\rangle
\langle {\bf B_3}|(\bar q_3 b)| {\bf B_b}\rangle$,
where $(\bar q_1 q_2)(\bar q_3 b)$ are
(axial)vector or (pseudo)scalar currents
from the quark-level effective Hamiltonian for charmless $b\to q_1\bar q_2 q_3$ transitions.
In the amplitude, the two matrix elements can be further presented as
the timelike baryonic form factors and the ${\bf B_b\to B_3}$ transition
form factors~\cite{Geng:2016fdw,Hsiao:2017nga,Hsiao:2018umx}, together with
the parameter for factorisable effects,
being decomposed as effective Wilson coefficients~\cite{Ali:1998eb},
the Fermi constant and Cabibbo-Kobayashi-Maskawa (CKM) matrix elements~\cite{CKM1,CKM2}.
The extractions of the form factors with their uncertainties
can be found in Refs.~\cite{Geng:2016fdw,Hsiao:2017nga,Hsiao:2018umx}.
The form factors have been used to calculate
${\cal B}(\bar B^0_s\to \Lambda \bar p K^+,\bar \Lambda p K^-)$~\cite{Geng:2016fdw},
whose value is in agreement with the measurement published by the LHCb collaboration~\cite{LHCb-PAPER-2017-012}.
Likewise, the prediction of the branching fraction ${\cal B}(\bar B^0\to \Lambda\bar p K^+ K^-)$
has been validated by the recently measurement by the \belle collaboration~\cite{Lu:2018qbw}.

Following the techniques described in previous work~\cite{Geng:2016drz},
the branching fractions for the three-body purely baryonic decays discussed
in this letter are predicted to be in the range $10^{-7}-10^{-6}$,
specifically
${\cal B}(\LbLPPbar) = (3.2 ^{+0.8}_{-0.3} \pm 0.4 \pm 0.7) \times 10^{-6}$ and
${\cal B}(\XibzLPPbar) =(1.4\pm 0.1\pm 0.1\pm 0.4)\times 10^{-7}$,
where the first uncertainties account for non-factorizable effects, the second reflect the
experimental knowledge of the CKM matrix elements, and the third arise from
those on the form factors~\cite{Geng:2016fdw,Hsiao:2017nga,Hsiao:2018umx}.

The direct \CP violating rate ($\Gamma$) asymmetry can be defined by
\begin{eqnarray}\label{acp1}
{\cal A}_{CP}=\frac{
\Gamma( {\bf B}_h\to{\bf B}_{l_1} \bar {\bf B}_{l_2} {\bf B}_{l_3})
-\Gamma( {\bf\bar B}_h\to{\bf \bar B}_{l_1}  {\bf  B}_{l_2} {\bf\bar B}_{l_3})}
{\Gamma( {\bf B}_h\to{\bf B}_{l_1} \bar {\bf B}_{l_2} {\bf B}_{l_3})
+\Gamma( {\bf\bar B}_h\to{\bf \bar B}_{l_1}  {\bf  B}_{l_2} {\bf\bar B}_{l_3})}\;.
\end{eqnarray}
If both  weak ($\gamma$) and strong ($\delta$) phases are non-vanishing, one has that
${\cal A}_{CP}\propto \sin\gamma\sin\delta$.
The direct \CP asymmetries of \LbLPPbar and \XibzLPPbar decays are predicted to be
$( 3.4 \pm 0.1 \pm 0.1 \pm 1.0 ) \%$ and $(-13.0\pm 0.5\pm 1.5\pm 1.1)\%$, respectively,
with the uncertainties mentioned early.

\addcontentsline{toc}{section}{References}
\setboolean{inbibliography}{true}
\ifx\mcitethebibliography\mciteundefinedmacro
\PackageError{LHCb.bst}{mciteplus.sty has not been loaded}
{This bibstyle requires the use of the mciteplus package.}\fi
\providecommand{\href}[2]{#2}

\section*{Acknowledgments}
The work of C.Q.~Geng and Y.K.~Hsiao was supported in part
by the National Center for Theoretical Sciences,
MoST (MoST-104-2112-M-007-003-MY3), and the
National Science Foundation of China (11675030).
The work of E. Rodrigues was supported in part
by the United States National Science Foundation award ACI-1450319.
E. R. wishes to thank the National Center for Theoretical Sciences
at Hsinchu, Taiwan, for its warm hospitality.

\section*{Author Contributions}
C.Q.G. and Y.K.H. performed the calculations and produced the figures.
All authors interpreted and analysed the results.
E.R. wrote the manuscript. All authors reviewed the manuscript.

\section*{Additional Information}
{\bf Competing Interests:} The authors declare no competing interests.

\end{document}